\documentclass{GR-QM-Direct}

\def\Journal#1#2#3#4{{#1} {\bf #2}, #3 (#4)}

\def\PLB{{\em Phys. Lett.}  B}
\def\PRL{\em Phys. Rev. Lett.}

\def\be{\begin{equation}}
\def\ee{\end{equation}}
\def\bea{\begin{eqnarray}}
\def\eea{\end{eqnarray}}

\newcommand{\Photo}{}

\begin{document}
\vspace*{4cm}

\title{Derivation of Klein-Gordon-Fock equation from General relativity in a time-space
symmetrical model}
\author{ VO VAN THUAN }
\address{Vietnam Atomic Energy Institute (VINATOM)
\\59 Ly Thuong Kiet street, Hoan Kiem district, Hanoi, Vietnam
\\Email: vvthuan@vinatom.gov.vn}
\maketitle\abstracts{
Following a bi-cylindrical model of geometrical dynamics, our current study has shown that
Einstein gravitational equation leads to bi-geodesic description in an extended symmetrical time-space which fits Hubble expansion in a "microscopic" cosmological model. As a duality, the geodesic solution is mathematically equivalent to the basic Klein-Gordon-Fock equations of free massive elementary particles, in particular, as the squared Dirac equations of leptons and as a sub-solution with pseudo-axion. This result would serve as an explicit approach to the consistency between quantum mechanics and general relativity. \\
Keywords: time-space symmetry, general relativity, microscopic cosmological model, wave-like solution, Klein-Gordon-Fock equation, axion.}
\section{Introduction}
Searching for the consistency between quantum mechanics and general relativity is the most important and long-standing issue of physics. Kaluza and Klein ~\cite{Ka1}$^{,}$~\cite{Kl1} were pioneers to propose a space-like extradimension (ED) which is compacted as a micro circle in relation to general relativity. For the semi-classical approach to quantum mechanics introduced by de Broglie and Bohm ~\cite{Br1}$^{,}$~\cite{Bo1} the hidden parameters are somehow reminiscent of EDs. However, the evidence for violation of Bell inequalities ~\cite{Be1}$^{,}$~\cite{Fr1} abandoned the models with local hidden parameters, leaving the door still open to non-local hidden variables. High dimensional superstring models have been developed following Kaluza-Klein geometrical dynamics, of which most applied space-like EDs, while few others considered time-like ones. There are two main approaches applying time-like EDs: membrane models in the Anti-de-Sitter geometry (AdS), such as ~\cite{Ma1}$^{,}$~\cite{Ra1} and induced matter models ~\cite{We1}$^{,}$~\cite{We2}. In particular, Maldacena ~\cite{Ma1} found a duality between AdS and conformal fields as AdS/CFT formalism. Randall and Sundrum ~\cite{Ra1} applied an infinite AdS 5D-model for a hierarchy solution. For the induced matter approach, Wesson ~\cite{We1} has proposed a space-time-matter model describing proper mass as a time-like ED. A geometrical dynamic model for
elementary particles was proposed by Koch ~\cite{Ko2} with a time-like ED which offered a method for derivation of Klein-Gordon equation. Following the induced-matter approach our preliminary study ~\cite{Vo1} was based on the time-space symmetry in which the Klein-Fock reduction formalism was used and the two time-like extra-dimensions were made explicit in terms of the quantum wave function $\psi$ and the proper time variable $t_0$. For a next step ~\cite{Vo2}, a duality was found between the quantum wave equations in 4D space-time and a relativistic geodesic description of the curved extradimensional time-space and, as a result, Heisenberg indeterminism is shown to originate from the space-time curvatures. Concerning the experimental verification, applying 3D-extended time-like curvatures can solve the problem of mass hierarchy of charge lepton generations ~\cite{Vo3} which opens a new possibility for extending to a more general solution of the heavy lepton-neutrino hierarchy. The goal of the present work is to look for a direct approach to the consistency between quantum mechanics and general relativity which would be implemented in particle physics. We are going to show that the extended general relativity equation leads to the desirable geodesic equation, which is mathematically identical to the basic Klein-Gordon-Fock quantum equation of free massive elementary particles. In particular, its content serves the basis for our conference talk on prediction of lepton hierarchy and absolute neutrino masses presented at the $51^{rst}$ Rencontre de Moriond-Cosmology, held in La Thuile, Italy, 19-26 March 2016 in memory of the centenary of Einstein general relativity and the $50^{th}$ anniversary of the Rencontres de Moriond.
\section{Time-space symmetrical geometry}
Considering the flat  $\{3T,3X\}$$\equiv$$\{t_1,t_2,t_3 \mid x_1,x_2,x_3\}$ symmetrical time-space:
\begin{equation}
dS^2=dt_k^2-dx_l^2,
\label{eq1}
\end{equation}
where $k,l=1 \div 3$ are summation indexes. Here we use mostly natural units throughout except when it needs  involving  quantum dimension.
The physics is investigated on the 6D-"lightcone" of time-space $(\ref{eq1})$:
\begin{equation}
dt_k^2=dx_l^2.
\label{eq2}
\end{equation}
In particular, considering in the lightcone manifold $(\ref{eq2})$ harmonic correlations $\psi_0[dt_k,dx_l]$ between a time-like and a space-like differentials-displacements as:
\begin{equation}
\frac{\partial^2 \psi_0 }{\partial t_k^2}=\frac{\partial^2 \psi_0 }{\partial x_l^2},
\label{eq3}
\end{equation}
Assuming that those homogenous and isotropic plane waves serve  a primitive source of quantum fluctuations in space-time.  All chaos of displacements $dt_k$  and $dx_l$ can form square-averaged time-like and space-like potentials $V_T$ and $V_X$, respectively. \\
For description of homogeneous and isotropic fluctuations  a spherical geometry is available. Suggesting that the global potentials, originally accelerating $\psi_0(t_k,x_l)$, turn plane waves into geodesic deviations with time-space symmetrical bi-spherical curvatures which are equivalent to isotropic spinning in symmetrical orthonormal subspaces of 3D-time and 3D-space. For elementary particles, except 3D-spatial spin $\vec s$, a pseudo-isospin $\vec \tau$ is introduced in 3D-time sphere. Therefore, the bi-spherical geometry with variables $\{\psi,\theta,\varphi\}$ reads:
\begin{equation}
{d\Sigma}^2=ds^2-{d\sigma}^2=dt^2-d\lambda^2 ,
\label{eq4}
\end{equation}
where: $dt^2=d\psi(t_k)^2+\psi(t_k)^2\left [ d\theta(t_k)^2+sin^2\theta(t_k) d\varphi(t_k)^2\right];$
\\and: $d\lambda^2=d\psi(x_l)^2+\psi(x_l)^2\left [ d\theta(x_l)^2+sin^2\theta(x_l) d\varphi(x_l)^2\right]$.\\

The original spherical coordinates $\{\psi,\theta,\varphi\}$  of Geometry $(\ref{eq4})$ being embedded in manifold $(\ref{eq2})$  are getting then functionally depending on linear coordinates $\{t_k,x_l\}$.
\\
For fluctuations including curved rotation and linear translation a $\{3D,3D\}$-symmetrical  bi-cylindrical geometry is applied:
\begin{equation}
d\Sigma^2=(ds_0^2+ds_{ev}^2)-(d\sigma_{ev}^2+d\sigma_L^2)=dt^2-d\lambda^2,
\label{eq5}
\end{equation}
where: $dt^2=d\psi(t_0,t_k)^2+\psi(t_0,t_k)^2d\varphi(t_0,t_k)^2+dt_k^2$
\\
and: $d\lambda^2=d\psi(x_n,x_l)^2+\psi(x_n,x_l)^2d\varphi(x_n,x_l)^2+dx_l^2.$
\\

In particular, observing  an individual fermion elementary particle, e.g.  a free lepton with (pseudo-)spins  $\vec \tau$ and $\vec s$  we can fix their projections $\tau_k$ or  $s_l=\pm 1/2$ on the longitudinal axes of $\{t_k\}$ and $\{x_l\}$, respectively, then: $d\theta^2=0$  and $sin^2\theta=1$. It leads to a cylindrical dynamical model. Embedding in Manifold $(\ref{eq2})$ the cylindrical variables $\psi$ and $\varphi$ are getting functions of linear coordinates $\{t_k,x_l\}$ and two 3D-local affine parameters $t_0$ and $x_n$ which are introduced in according to projection of (pseudo-)spins  $\vec \tau$ and $\vec s$, respectively, namely:
 \begin{equation}
\psi=\psi(t_0,t_k,x_n,x_l) ; \qquad
\varphi=\Omega_0t_0+\Omega_kt_k-k_n x_n-k_l x_l=\Omega_i t_i-k_jx_j ,
\label{eq6}
\end{equation}
where $\{i,j\}$ are summation indexes of curved coordinates. For an explicit description, $\{t_3, x_3\}$ are accepted in Geometry $(\ref{eq5})$ as longitudinal central axes of the bi-cylinder, the curved coordinates of 3D-space are $\{x_j\}\in \{x_1,x_2,z\}$ with $k^2dz^2=k_n^2dx_n^2+k_3^2dx_3^2$. Similarly, for 3D-time there are $\{t_i \}\in \{t_1,t_2,t\}$  where the longitudinal axis $t_3$ can combine with the rotational affine parameter $t_0$ to form the real physical time $t$ by an orthogonal relationship: $\Omega^2dt^2=\Omega_0^2dt_0^2+\Omega_3^2dt_3^2$.
\\
In more general case of dynamical fluctuations the cylindrical  and spherical geometries are combined, where spherical coordinates are reserved for describing a potential  even-contribution from 3D-rotation of  pseudo-spin or spin. In the following scenario due to interaction of  a Higgs-like potential the time-space symmetry is  spontaneously broken, leading to formation of energy-momentum,  which makes the spherical curvature turned to an almost exact  cylindrical curvature. In the result, Geometry $(\ref{eq5})$ turns to an asymmetrical bi-cylindrical geometry:
\begin{equation}
d\Sigma^2=(ds_0^2+ds_{ev}^2)-(d\sigma_{ev}^2+d\sigma_L^2)=dt^2-dz^2,
\label{eq5a}
\end{equation}
where: $dt^2=d\psi(t_0)^2+\psi(t_0)^2d\varphi(t_0)^2+dt_3^2$  and: $dz^2=d\psi(x_n)^2+\psi(x_n)^2d\varphi(x_n)^2+dx_3^2.$
\\

The time-like and space-like intervals in Geometry $(\ref{eq5a})$ separate into even and odd constituents, which means the corresponding cylindrical accelerations can flip for and back  (as an even-term) or can not flip (as an odd-term)  in relation to the cylindrical axis. Odd-terms are determined by internal curvatures, while even-terms relate to external curvatures of cylinders embedded in corresponding flat 3D-subspaces. Naturally, an "internal observer" involved in a spiral rotation, being not able to distinguish this cylindrical curvature, considers its space-time flat. In a semi-phenomenological consideration, the T-odd term $ds_0$ is equivalent to a conventional interval  of special relativity in 4D-Minkowski space-time, while the P-odd term $d\sigma_L$ is a P-nonconserving contribution (PNC) of the weak interaction making a global space-like curvature in term of the left-handed helicity of fermion elementary particles. The P-even term $d\sigma_{ev}$ describes 3D-space kinetics then $x_3$ is selected as a cylindrical axis, implying that  $x_3\in \{x_l\}$. In principle, a free electron reserves its angular momentum (spin) correlating with fixed $d\sigma_{ev}$, but in an external laboratory frame because the proper angular momentum is no more observable unless using an appropriate on-line polarization analysis, then  $d\sigma_{ev}$ is getting hidden by a geodesic compensation. In the result, all three linear spatial axes $\{x_l\}$ reveals in transformation from the cylindrical frame to 3D-space laboratory frame. In 4D-Minkowski space-time the P-odd term is too small as $d\sigma_L\ll d\sigma_{ev}$, that makes the weakly curved $\{x_j\}\approx\{x_l\}$ which open as it is observable in a flat 3D-space. The time-like even-term $ds_{ev}\equiv d\sigma_{CPV}$  is introduced to describe CP-violation effects in 3D-time.\\
Therefore, an asymmetry of Geometry $(\ref{eq5a})$ means that in 4D-Minkowski subluminal space-time $d\sigma_{CPV}\ll d\sigma_L\ll d\sigma_{ev}\ll ds_0$, that $d\sigma_{CPV}$  and $d\sigma_L$ may be ignored. In the result, the 3D-time curvature is getting absolute and the subluminal physics is involved in evolution along the physical time $t$, in the meantime, $d\sigma_{ev}$ turns to a pseudo-cylindrical interval in an almost linear 3D-space $\{x_j \}$ with a weakly curved residue along axis $z$. The curvatures are described by EDs $\psi$ and $\varphi$ in $(\ref{eq6})$. The 4D-subluminal geometry fits the charged lepton sector ~\cite{Vo3}.\\
For 4D-superluminal time-space, it is assumed that $d\sigma_{ev}\ll ds_0\ll d\sigma_{CPV}\ll d\sigma_L$ then 3D-time is almost flat, while $d\sigma_{ev}$  and $ds_0$ may be ignored. In the result, the 3D-spatial curvature is getting absolute and superluminal substances  are to move along the physical 3D-curved $z$; in the meantime, for subluminal observer $d\sigma_{CPV} \ll d\sigma_L$ then  $d\sigma_{CPV}$ turns to a pseudo-cylindrical interval in an almost flat 3D-time. Such a strong asymmetry between space-like and time-like curvatures, as well as a possibility of exchanging the roles of space $\Leftrightarrow$ time (or equivalently, momentum $\Leftrightarrow$ energy) would be able to fit neutrino sector, in particular, for solving the problem of large hierarchy between neutrinos and heavy leptons.
\section{Wave-like solutions of Einstein gravitational equation}
Suggesting that in both orthonormal subspaces of 3D-time and 3D-space cylindrical curvature is realized. The gravitational equation in  $\{3T,3X\}$-vacuum (in absence of matter) of Geometry $(\ref{eq5})$ reads:
\begin{equation}
R^m_i-\frac{1}{2}\delta^m_i R=-\delta^m_i\Lambda,
\label{eq7}
\end{equation}
where $R^m_i$ and $R$ are tensor and scalar curvatures, respectively. When cosmological constant is vanished  ($\Lambda=0$) the $\{3T,3X\}$ equation  $(\ref{eq7})$  is separated into two independent 3D sub-equations. It leads to a time-space symmetrical representation:
\begin{equation}
R_\alpha^\beta (T)-\frac{1}{2} \delta^\beta_\alpha R(T)=R_\gamma^\sigma (X)-\frac{1}{2} \delta^\sigma_\gamma R(X),
\label{eq8}
\end{equation}
where tensors with $\alpha,\beta\in$ 3D-time  and ones with $\gamma,\sigma\in$ 3D-space.
\\Calculating related Christoffel symbols, as $\psi=\psi(y)$ and  $\varphi=\varphi(y)$ are being not independent variables, we assume that the Hubble law of the cosmological expansion is able to be applied for the bi-cylindrical model of microscopic space-time:
$\frac{\partial \psi}{\partial y} =v_y=H_y\psi$. Therefore:
\begin{equation}
\left[\frac{\partial y}{\partial \psi}\right]=\frac{1}{H_y \psi},
\label{eq9}
\end{equation}
where the expansion rate $v_y$ increases proportional to  the "microscopic scale factor" $\psi$ , and $H_y$  is the "microscopic Hubble constant".
The following Christoffel symbols are found valid:\\
\begin{equation}
\Gamma_{\varphi\varphi}^\psi=-\frac{g^{\psi\psi}}{2}\frac{\partial g_{\varphi\varphi}}{\partial\psi}=-\frac{1}{H_y}\frac{\partial\psi}{\partial y};
\label{eqK1}
\end{equation}
\begin{equation}
\Gamma_{\psi\varphi}^\varphi=\Gamma_{\varphi\psi}^\varphi=\frac{g^{\varphi\varphi}}{2}\frac{\partial g_{\varphi\varphi}}{\partial\psi}=\frac{1}{\psi^2H_y}\frac{\partial\psi}{\partial y};
\label{eqK2}
\end{equation}
\begin{equation}
\Gamma_{\varphi\varphi}^3=-\frac{g^{33}}{2}\frac{\partial g_{\varphi\varphi}}{\partial y_3}=-\psi \frac{\partial\psi}{\partial y_3};
\label{eqK3}
\end{equation}
\begin{equation}
\Gamma_{3\varphi}^\varphi=\Gamma_{\varphi 3}^\varphi=\frac{g^{\varphi\varphi}}{2}\frac{\partial g_{\varphi\varphi}}{\partial y_3}=\frac{1}{\psi}\frac{\partial\psi}{\partial y_3},
\label{eqK4}
\end{equation}
where real space-time variables $y\equiv\{t,z\}\equiv\{t_0,t_3,x_n,x_3\}\in \{t_i,x_j\}$; and $y_3\equiv\{t_3,x_3\}\in \{t_k,x_l\}$ as being implicitly embedded in 3D-time and in 3D-space, respectively.\\
The tensor curvatures are calculated for the 3D-local cylindrical geometry as:\\
\begin{equation}
R_{33}=-\frac{1}{\psi}\frac{\partial^2\psi}{\partial y^2_3};
\label{eqR1}
\end{equation}
\begin{equation}
R_{\psi\psi}=-\frac{1}{\psi^3 H^2_y}\frac{\partial^2\psi}{\partial y^2}+\frac{1}{\psi^4 H^2_y} \left (\frac{\partial\psi}{\partial y} \right )^2;
\label{eqR2}
\end{equation}
\begin{equation}
R_{\varphi\varphi}=-\frac{1}{\psi H^2_y}\frac{\partial^2\psi}{\partial y^2}+\frac{1}{\psi^2 H^2_y} \left (\frac{\partial\psi}{\partial y}\right )^2-\psi \frac{\partial^2\psi}{\partial y^2_3}.
\label{eqR3}
\end{equation}
Obviously, scalar curvature reads:
\begin{equation}
R=g^{im}R_{im}=R_{\psi \psi}+\psi^{-2} R_{\varphi \varphi}+R_{33}=R^i_i=-\frac{2}{\psi^3 H^2_y}\frac{\partial^2\psi}{\partial y^2}+\frac{2}{\psi^4 H^2_y} \left (\frac{\partial\psi}{\partial y}\right )^2-\frac{2}{\psi}\frac{\partial^2\psi}{\partial y^2_3}.
\label{eqR}
\end{equation}
Its time-space separable form reads: $R=\delta^\sigma_\gamma R^\sigma_\gamma(X)-\delta^\beta_\alpha R^\beta_\alpha(T)$.
\\
 It leads to the gravitational equation in $\{3T,3X\}$-vacuum in absence of matter, but with the cosmological constant as a kind of global
deviation between local curvatures in 3D-time and 3D-space, namely: $\Lambda\equiv\{\Lambda_{\psi}^{\psi},\Lambda_{\varphi}^{\varphi},\Lambda_3^3\}$. \\
Therefore, Equation  $(\ref{eq7})$  is equivalent to a system of component equations:
\begin{equation}
R_3^3=\Lambda_3^3;
\label{eq10}
\end{equation}
\begin{equation}
R^\psi_\psi=\Lambda_{\psi}^{\psi};
\label{eq11}
\end{equation}
\begin{equation}
R^\varphi_\varphi=\Lambda_{\varphi}^{\varphi}.
\label{eq12}
\end{equation}
Equation $(\ref{eq12})$ is a trivial relation which is clear from:  $R^\varphi_\varphi=R^\psi_\psi+R_3^3 $, then it means: $\Lambda_{\varphi}^{\varphi}=\Lambda_{\psi}^{\psi}+\Lambda_3^3$. For the homogeneous and isotropic space-time it implies that $\Lambda_{\psi}^{\psi}=\Lambda_3^3$.
\\
There only two sub-equations $(\ref{eq10})$ and $(\ref{eq11})$ are independent.\\
Equation $(\ref{eq10})$ in details reads:
\begin{equation}
R_3^3=-\frac{1}{\psi}\frac{\partial^2\psi}{\partial y^2_3}=R_3^3 (X)-R_3^3 (T)=\Lambda_3^3.
\label{eq13}
\end{equation}
Originally, Equation $(\ref{eq13})$ is almost time-space symmetrical except small contributions of $\Lambda_3^3$, then it leads to a formula:
\begin{equation}
-\frac{\partial^2 \psi}{\partial t_3^2}+\frac{\partial^2 \psi}{\partial x_3^2}\equiv -\frac{\partial^2 \psi}{\partial t^2}+\frac{\partial^2 \psi}{\partial x_j^2}=-\Lambda_3^3 \psi.
\label{eq13a}
\end{equation}
For $\Lambda_3^3<0$, rescaling $(\ref{eq13a})$ with quantum Planck constant, we obtain Klein-Gordon-Fock equation with a pseudo-plane wave-like solution  describing a scalar particle of a tiny mass $m_a\sim \sqrt {-\Lambda_3^3}$:
\begin{equation}
-\hbar^2\frac{\partial^2 \psi}{\partial t^2}+\hbar^2\frac{\partial^2 \psi}{\partial x_j^2}=m_a^2\psi.
\label{eq13b}
\end{equation}
Because $\Lambda_3^3$, as a parameter governing time-space asymmetry, including a contribution of longitudinal fluctuations violating Lorentz invariance, then it would naturally lead to CP violation. Such a pseudo-particle described by Equation $(\ref{eq13b})$ may fit axion, which was predicted in Peccei-Quinn model ~\cite{Qu1} to maintain a strong CPV in the universe.\\
When $\Lambda=0$ from $(\ref{eq13})$ we obtain a plane wave equation:
\begin{equation}
-\frac{\partial^2 \psi}{\partial t_3^2}+\frac{\partial^2 \psi}{\partial x_3^2}=0.
\label{eq14}
\end{equation}
Indeed, Equation $(\ref{eq14})$ has a wave-like solution $\psi=\psi_0 e^{\mp i(\omega_3 t_3-k_3 x_3)}$ which leads to an additional Lorentz-like condition:
\begin{equation}
(\omega_3^2-k_3^2)\psi=0.
\label{eq15}
\end{equation}
which compensates any longitudinal fluctuations. Therefore, Equation $(\ref{eq14})$ defines conservation of  linear translation (CLT) which originates from Equation  $(\ref{eq3})$. \\
Equation $(\ref{eq11})$ determines a bi-geodesic equation:
\begin{equation}
R_\psi^\psi=-\frac{1}{\psi^3H_y^2} \frac{\partial^2 \psi}{\partial y^2}+\frac{1}{\psi^4H_y^2}\left (\frac{\partial \psi}{\partial y}\right )\psi=R_\psi^\psi (X)-R_\psi^\psi (T)=\Lambda_{\psi}^{\psi}.
\label{eq16}
\end{equation}
It leads to an original symmetrical equation:
\begin{equation}
 \frac{\partial^2 \psi}{\partial y^2}-\left (\frac{\partial \varphi}{\partial y}\right )^2(1-\Lambda_{\psi}^{\psi}\psi^2)\psi=0.
\label{eq16a}
\end{equation}
Naturally, $\Lambda_{\psi}^{\psi}\psi^2\ll 1$, then for real $\{y\}$ Equation $(\ref{eq16a})$ having an exponential solution $\psi\sim e^{H_y.y}=e^\varphi=e^{\Omega t-k_j x_j}$ is getting:
\begin{equation}
\frac{\partial^2\psi}{\partial t^2}-\left[\left(\frac{\partial \varphi}{\partial t_0}\right)^2+\left(\frac{\partial \varphi}{\partial t_3}\right)^2\right]\psi=\frac{\partial^2\psi}{\partial x_j^2}-\left[\left(\frac{\partial \varphi}{\partial x_n}\right)^2+\left(\frac{\partial \varphi}{\partial x_l}\right)^2\right]\psi.
\label{eq17}
\end{equation}
As differentials $dt_3$ and $dt_0$, as well as corresponding covariant derivatives are locally orthogonal to each other, their second derivatives are combined together as: $\frac{\partial^2 \psi}{\partial t^2}=\frac{\partial^2 \psi}{\partial {t_0}^2}+\frac{\partial^2 \psi}{\partial {t_3}^2}$.
\\
Similarly, due to a local orthogonality, for differentials $dx_l$ and $dx_n$, the second derivatives in 3D-space are also combined:
$\frac{\partial^2 \psi}{\partial {x_j}^2}=\frac{\partial^2 \psi}{\partial {x_n}^2}+\frac{\partial^2 \psi}{\partial {x_l}^2}$.
\\
For a homogeneity condition, Equation $(\ref{eq17})$ is getting a symmetrical equation of bi-geodesic acceleration of deviation $\psi$ in 3D-time and 3D-space as obtained in ~\cite{Vo2}:
\begin{equation}
\frac{\partial^2\psi}{\partial t_0^2}-\left(\frac{\partial \varphi}{\partial t_0}\right)^2\psi=\frac{\partial^2\psi}{\partial x_n^2}-\left(\frac{\partial \varphi}{\partial x_n}\right)^2\psi.
\label{eq18}
\end{equation}
Recalling that due to 3D local geodesic condition when $\Lambda=0$, both sides in $(\ref{eq18})$ are getting independent and lead to de Sitter-like exponential sub-solutions  which describe Hubble-like expansion in microscopic local 3D-time or local 3D-space, correspondingly:
\begin{equation}
\frac{\partial^2\psi}{\partial t_0^2}-\left(\frac{\partial \varphi}{\partial t_0}\right)^2\psi\Rightarrow\frac{\partial^2\psi}{\partial {t_0^+}^2}-\left(\frac{\partial \varphi}{\partial t_0^+}\right)^2\psi=\frac{\partial^2\psi}{\partial {t_0^+}^2}-\Lambda_T\psi=0;
\label{eq18T}
\end{equation}
\begin{equation}
\frac{\partial^2\psi}{\partial x_n^2}-\left(\frac{\partial \varphi}{\partial x_n}\right)^2\psi\Rightarrow\frac{\partial^2\psi}{\partial {x_n}^2}-\left(\frac{\partial \varphi}{\partial x_n}\right)^2\psi=\frac{\partial^2\psi}{\partial {x_n}^2}-\Lambda_X\psi=0,
\label{eq18X}
\end{equation}
where effective strong potentials $V_T$ of a time-like "cosmological constant" $\Lambda_T$ with a minor space-like component $\Lambda_X$ fulfill breaking symmetry, making time oriented toward the future ($t_0^+$-signature) and space getting longitudinal polarization ($x_n$-signature of a fixed P-even spin projection). In ~\cite{Vo3} we suggested that Equation $(\ref{eq18T})$ can fit a scenario similar to the standard cosmological model to formulate a so-called microscopic cosmological model which is able to predict comoving volumes, then calculating masses of charged leptons. Therefore, the mass hierarchy problem of leptons would be solved, based on time-space symmetry.
\\
Equation $(\ref{eq17})$ with the CLT principle and Lorentz-like condition $(\ref{eq15})$  leads to the symmetrical exponential geodesic solution in according to Geometry  $(\ref{eq5})$ as following:
\begin{equation}
-\frac{\partial^2\psi}{\partial t^2}+\frac{\partial^2\psi}{\partial x_j^2}=-\left[\left(\frac{\partial \varphi}{\partial t_0}\right)^2-\left(\frac{\partial \varphi}{\partial x_n}\right)^2 \right]\psi.
\label{eq17a}
\end{equation}
\\
In principle, variables  $\{y\}$ can turn as well as $\{y\}\leftrightarrow \{iy\}$  in a mathematical transformation, then Condition $(\ref{eq9})$ turns to:
\begin{equation}
\left[\frac{\partial y}{\partial \psi}\right]=\frac{-i}{H_y \psi}.
\label{eq9a}
\end{equation}
In the result, Equation $(\ref{eq17a})$ leads to another representation of a wave-like solution with $\psi(y\rightarrow iy)\sim e^{i\varphi}=e^{i(\Omega t-k_j x_j)}$ as following:
\begin{equation}
-\frac{\partial^2\psi}{\partial t^2}+\frac{\partial^2\psi}{\partial x_j^2}=\left[\left(\frac{\partial \varphi}{\partial t_0}\right)^2-\left(\frac{\partial \varphi}{\partial x_n}\right)^2 \right]\psi.
\label{eq19}
\end{equation}
If time-space symmetry is absolute, the right side is vanished and Equation $(\ref{eq19})$ turns to:
\begin{equation}
-\frac{\partial^2\psi}{\partial t^2}+\frac{\partial^2\psi}{\partial x_j^2}=0.
\label{eq19a}
\end{equation}
Being involved in metrics $g_{\varphi\varphi}$ the functional parameter $\psi$  characterizes time-space  curvatures. Then Equation $(\ref{eq19a})$  would describe a specific kind of microscopic gravitational waves transmitting with the speed of light.
However, the time-space symmetry can never be absolute: we have already assumed that the acceleration term in 3D-time  is dominantly enhanced due to interaction with a Higgs-like potential, that will produce a time-space asymmetrical polarization $P\rightarrow P^+$.
\\
Qualitatively, the original $\{3T,3X\}$ time-space symmetry is broken spontaneously:
\begin{equation}
(V_TP)^2=\left [V_T \left (\frac{\partial \varphi}{\partial t_0^-}+\frac{\partial \varphi}{\partial t_0^+} \right) \right ]^2 \psi \equiv [f_e(\chi+\phi_0)]^2\psi\Rightarrow (P^+)^2=\left (\frac{\partial \varphi}{\partial t_0^+} \right)^2\psi\equiv (f_e\phi_0)^2\psi=m_0^2 \psi,
\label{eq20}
\end{equation}
where $\chi$ is Higgs field and $\phi_0$ is Higgs vacuum; $f_e$ is Higgs-lepton coupling constant. The arrow means the moment of fixing polarization, equivalent to a spontaneous breaking of symmetry. Since that the elementary particle as a material point has been involved in an almost absolute time-like cylindrical evolution along the real time $t$. Any human observation along the same local geodesic in 3D-time can not distinguish any spiral evolution because the internal curvature of a cylinder is zero. This is the reason explaining why the physical time axis of a freely moving elementary particle is linear in 4D-Minkowski space-time.
\\
After spontaneous breaking of time-space symmetry, Equation $(\ref{eq17a})$ determines an asymmetrical bi-geodesic equation with exponential solutions in according to Geometry  $(\ref{eq5a})$:
\begin{equation}
\frac{\partial^2 \psi}{\partial t^2}-\frac{\partial^2 \psi}{\partial {x_j}^2}=\left [\Lambda_T -\left(\frac{\partial \varphi}{\partial x_n}\right)_{even}^2 -\Lambda_L \right ]\psi.
\label{eq21}
\end{equation}
where $\Lambda_L\equiv \left(\frac{\partial \varphi}{\partial x_n^L}\right)^2$ is a small space-like P-odd "cosmological constant" caused by the global weak interaction leading to the left-handed space. Being originated from Einstein gravitational equation $(\ref{eq7})$, Equation $(\ref{eq21})$ describes the microscopic cosmological evolution of time-space curvatures by its de Sitter-like exponential solutions $\psi=\psi_0 e^{\pm \varphi}=\psi_0 e^{\pm (\Omega t+k_j x_j)}$.
\\
Correspondingly, the wave-like equation $(\ref{eq19})$ with $\psi_w(y)\equiv\psi(iy)=\psi_0 e^{\pm i\varphi}=\psi_0 e^{\mp i(\Omega t-k_j x_j)}$  due to breaking  symmetry leads to:
\begin{equation}
-\frac{\partial^2 \psi}{\partial t^2}+\frac{\partial^2 \psi}{\partial {x_j}^2}=\left [\left(\frac{\partial \varphi}{\partial t_0^+}\right)^2-B_e ( k_n.\mu_e)_{even}^2 -\left(\frac{\partial \varphi}{\partial x_n^L}\right)^2 \right ]\psi.
\label{eq22}
\end{equation}
where $B_e$ is a calibration factor  and $\mu_e$ is magnetic dipole moment of charged lepton;  its orientation is in correlation with spin vector $\vec s$ and being P-even.
\section{The duality of general relativity for interpretation of quantum mechanics}
In mathematical transformation from the exponential solution to the wave-like one, we should change the signature in Equation $(\ref{eq21})$, keeping the wave equation $(\ref{eq22})$ mathematically equivalent to the former. This is realized by  transformation of variables: $t\rightarrow -it$ and $x_j \rightarrow ix_j$, as well as of their corresponding covariant derivatives: $\frac{\partial f}{\partial t}\rightarrow i\frac{\partial f}{\partial t}$ and $\frac{\partial f}{\partial x_j} \rightarrow -i\frac{\partial f}{\partial x_j}$, similarly as being adopted  for quantum dynamic operators. This procedure is not only a mathematical formalism, but also a significant physical operation, equivalent to transformation from an external observation to an internal investigation.  Really, it is a fact in quantum mechanics that the phase velocity in the internal phase continuum is superluminal as clear from $\varphi=\Omega t-k_j x_j=const$, then $v_{phase}=\frac{dx_j}{dt}=\frac{\Omega}{k_j}=\frac{E}{p_j} > c$. Somehow, it is equivalent to converting the role of space $\Leftrightarrow$ time in the internal superluminal frame comparing with the external subluminal space-time. Indeed, instead of the real time in 4D-Minkowski geometry one can use an imaginary time in the corresponding 4D pseudo-Euclid representation. The latter with space-time $\{x,-it\}$ is explicitly symmetrical in a mathematical transformation $\{x,-it\} \Leftrightarrow\{ix,t\}$, but for an observation in the subluminal frame the imaginary coordinate is equivalent to a time axis, while the "real time" can be accepted as a spatial axis. Therefore, by rescaling dynamic action with Planck constant (namely, implementing quantum dynamical operators $\frac{\partial }{\partial t}\rightarrow \hat{E}=i\hbar\frac{\partial}{\partial t}$ and $\frac{\partial }{\partial x_j} \rightarrow \hat{p_j}=-i\hbar\frac{\partial }{\partial x_j}$) and making the amplitude of the functional parameter $\psi$ of a scale of Compton length, Klein-Gordon-Fock equation in quantum mechanics is to be formulated explicitly from the wave-like solution $(\ref{eq22})$ of Einstein gravitational equation $(\ref{eq7})$ in the extended time-space as:
\begin{equation}
-\hbar^2\frac{\partial^2 \psi}{\partial t^2}+\hbar^2\frac{\partial^2 \psi}{\partial x_j^2}-m^2\psi=0,
\label{eq23}
\end{equation}
where the square mass term $m$ consists of the following components:   $m^2=[\hbar\Omega]^2=m_0^2-\delta m^2=m_0^2-m_s^2-m_L^2$. Applying Fourier transformation from the phasic space to momentum representation, Equation $(\ref{eq23})$ reads:
\begin{equation}
E^2\psi_p-\vec{p}^2\psi_p-m^2\psi_p=0.
\label{eq24}
\end{equation}
It leads to the relation $E^2-\vec{p}^2=m^2>0$, then Equation $(\ref{eq24})$ describes subluminal motion of an elementary particle with energy $E$ and momentum $\vec{p}$. In comparison with the traditional expression of the rest mass, the present one includes an additional correction $\delta m$ associated with the contribution of the intrinsic spin in 3D-space. The P-even contribution $m_s$ linked with an external curvature of spinning in 3D-space can be compensated in according to 3D-spatial local geodesic condition in Equations $(\ref{eq21})$ and $(\ref{eq18X})$ when only the linear translation along $x_l$ axis is taken in account for a laboratory frame observation. However, due to P-odd effect being observable in the weak interaction, the geodesic deviation of the material point by its spinning still induces a small non-zero mass factor $m_L\ll m_s$ which proves a tiny internal curvature of our realistic 3D-space. The latter, in similar to the time-like cylindrical curvature of real time $t$, is not observable from the human point of view of 4D space-time observers, being involved as well in the same global internal cylindrical curvature caused by the weak interaction. In general, Equation $(\ref{eq23})$ is reminiscent of the squared Dirac equation of lepton~\cite{Vo1}. In case when there is no polarization analysis, $m \rightarrow m_0$, Equation $(\ref{eq23})$ turns to the traditional Klein-Gordon-Fock equation in the linear 3D-space (with $\{ {x_j} \} \rightarrow \{ {x_l} \}$):
\begin{equation}
-\hbar^2\frac{\partial^2 \psi}{\partial t^2}+\hbar^2\frac{\partial^2 \psi}{\partial x_l^2}-{m_0}^2\psi=0.
\label{eq25}
\end{equation}
The consistency between the physical reality of an individual elementary particle and the quantum statistical interpretation is a dilemma causing an unsolved philosophical problem. Our proposed model of bi-cylindrical geometrical dynamics would contribute to understanding some issues of this matter. First of all, it gives a meaning of the traditional quantum dynamical operators as time-space converting transformation in the phasic continuum, where the phase velocity of a massive particle is always faster than light. The specific kind of gravitational waves carrying the functional variable $\psi$ of  metrics $g_{\varphi \varphi}$ along with in the phasic continuum should be superluminal in this context, and it would be a reason why ones can not observe directly the quantum wave function $\psi$ except its squared amplitude.
\\Secondly, it would shed light on the wave-particle duality of quantum mechanics. Qualitatively, the wave-like sub-equation $(\ref{eq22})$ leads to description of a quantum substance as a non-local object with quantum wave features in 4D space-time. On other side, the exponential solution $(\ref{eq21})$ dually describes the same object, but  as a material point, i.e. a localized particle following a classical geodesics in an extended time-space.
\\Thirdly, from the homogeneous $\{3T,3X\}$ bi-geodesic condition $(\ref{eq18})$ it is possible to derive Heisenberg inequalities from space-time curvatures as shown in  ~\cite{Vo2}. Indeed, Equations $(\ref{eq18T})$ and $(\ref{eq18X})$ lead to the following local relationships in 3D-time and 3D-space, respectively:
\begin{equation}
d E_0.d t_0^+=\psi^{-1}d\left ( i.\hbar\frac{\partial \psi}{\partial t_0^+} \right )dt_0^+=i.\hbar.d\varphi^2.
\label{eq31a}
\end{equation}
\begin{equation}
d p_n.d x_n= \psi^{-1}d\left ( -i.\hbar\frac{\partial \psi}{\partial x_n} \right )dx_n=-i.\hbar.d\varphi^2.
\label{eq32a}
\end{equation}
Based on Equation $(\ref{eq31a})$ the time-energy inequality is derived:
\begin{equation}
\left | \Delta E \right |.\left | \Delta t \right |\geq \left | \Delta E_0 \right |.\left | \Delta t_0 \right |>\left |d E_0\right |.\left |d t_0^+\right |=\left | i.\hbar \right |.d\varphi^2\geq \Delta \varphi_{min}^2\hbar\geq{0}.
\label{eq43}
\end{equation}
Similarly, applying Equation $(\ref{eq32a})$ leads to the space-momentum inequality:
\begin{equation}
\left | \Delta p \right |.\left | \Delta x \right |\geq \left | \Delta p_n \right |.\left | \Delta x_n \right |> \left |d p_n\right |.\left |d x_n\right |=\left | i.\hbar \right |. d\varphi^2\geq \Delta \varphi_{min}^2\hbar\geq{0}.
\label{eq44}
\end{equation}
Obviously, the acceleration terms in Equations $(\ref{eq18T})$ and $(\ref{eq18X})$ would be vanished to turn the inequalities in $(\ref{eq43})$ and $(\ref{eq44})$ equal to zero only when space-time is getting flat. For a non-zero curvature, there is assumed that both inequalities adopt the condition $\Delta \varphi_{min}=\sigma\left(<\varphi>\right)=\sqrt{2\pi}$, where due to a statistical observability of the quantum indeterminism, $\sigma$ is a standard deviation of the mean value $<\varphi>=2\pi$ as a minimum of eigenvalues of the quantized azimuthal angle ($\varphi=2n\pi$) in an appropriate stochastic statistical distribution (Poisson or Gaussian one). The statistical feature is possibly caused by collective effects of non-localized interaction in 4D space-time between a detector system consisted of both human action and experimental apparatus with the observable microscopic object, however, being distorted in unpredicted way as an individual elementary particle. In the result, this leads to the traditional Heisenberg inequalities:
\begin{equation}
\left | \Delta E \right |.\left | \Delta t \right |\geq 2\pi\hbar.
\label{eq43a}
\end{equation}
\begin{equation}
\left | \Delta p \right |.\left | \Delta x \right |\geq 2\pi\hbar.
\label{eq44a}
\end{equation}
Therefore, the quantum indeterminism is found to originate from time-like and space-like curvatures, namely, the time-energy inequality is caused by an intrinsic curvature of 3D-time, while the space-momentum inequality is caused by a P-even contribution of spinning in 3D-space of an individual elementary particle.
\section{Conclusion}
Based on time-space symmetry a geometrical dynamical model was developed with an intrinsic cylindrical curvature determined by the functional parameter $\psi$.  According to this geometry, Einstein gravitational equation in vacuum has a duality: an exponential solution and another wave-like representation. As it is able to derive Klein-Gordon-Fock equation of a massive elementary particle from the wave-like solution, it was found that quantum mechanics is a special technique with its quantum dynamical operators for describing the superluminal microscopic gravitational waves in the microscopic phase continuum carrying energy-momentum in the corresponding subluminal macroscopic space-time. In addition, a solution with quasi-plane waves fits an assumption of a light scalar pseudo-particle, being reminiscent of axion.  On other side, the exponential solution leads to a geodesic description of an elementary particle, in particular a massive lepton, as a material point in the extended time-space. In combination, the dual solutions could shed light on origin of the quantum indeterminism and other important issues of quantum mechanics, such as the wave-particle duality. In a homogeneity condition the geodesic equation being equivalent to de Sitter-like solutions can serve for modeling  Hubble expansion in the microscopic time-space in analogue to the standard model of macroscopic cosmology. In particular, the proposed microscopic cosmological model with an extension of time-like EDs to the 3D-time formalism which correlates strictly with the number three of lepton generations, has been used to solve the mass hierarchy problem of charged leptons ~\cite{Vo3}. As a next step, the time-space symmetry and its breaking are expected to extend for clarifying the large hierarchy between heavy lepton and neutrino sectors, then a prediction of neutrino absolute masses would be an experimental challenge for verification of the model. In principle, neutrinos are very promising messengers for understanding the objective physical reality of elementary particles. Consequently, findings from time-space symmetry of microscopic substances would demonstrate a deep consistency between quantum mechanics and general relativity.
\section*{Acknowledgment}
The author is grateful to Dr. Tran Chi Thanh (Vietnam Atomic Energy Institute) for VINATOM support. We thank Nguyen Anh Ky (Hanoi Institute of Physics) and Do Quoc Tuan (Physics Faculty, Hanoi University of Natural Sciences) for useful discussions. A heart-felt thank is also extended to N.B. Nguyen (Thang Long University) for valuable technical assistance.

\end{document}